\RequirePackage{fixltx2e}

\documentclass[aip,pof,reprint]{revtex4-1}
\usepackage{graphicx,bm,amsmath,enumerate,amssymb,amsthm,color}

\begin{document}


\title{A volume-of-fluid formulation for the study of co-flowing fluids 
governed by the Hele-Shaw equations}



\author{Shahriar Afkhami}
\email[]{shahriar.afkhami@njit.edu}
\affiliation{Department of Mathematical Sciences, University Heights, New Jersey Institute of 
Technology, Newark, NJ 07102-1982, USA}
\author{Yuriko Renardy}
\affiliation{Department of Mathematics, 460 McBryde Hall, Virginia Tech, Blacksburg VA 24061-0123, 
USA}

\date{\today}
\begin{abstract}
We present a computational framework to address the flow of two immiscible viscous
liquids which co-flow into a shallow rectangular container at one side,
and flow out into a holding container at the opposite side. Assumptions
based on the shallow depth of the domain are used to reduce the governing
equations to one of Hele-Shaw type.  The distinctive feature of the numerical
method is the accurate modeling of the capillary effects. A continuum approach coupled with a 
volume-of-fluid formulation for computing the interface motion and for modeling
the interfacial tension in Hele-Shaw flows are formulated
and implemented. The interface is reconstructed with a height-function 
algorithm. The combination of these algorithms is a novel development for
the investigation of Hele-Shaw flows. The order of accuracy and convergence
properties of the method are discussed with benchmark simulations.
 A microfluidic flow of a  ribbon of fluid
which co-flows with a second liquid is  simulated.  We show that for small capillary numbers of
O(0.01), there is an abrupt change in interface curvature and focusing occurs close to the exit.
\end{abstract}



\maketitle 


\section{\label{sec:background}Introduction}
Microfluidic devices for droplet production are often based on forcing a jet of
one liquid sandwiched in another liquid  through  a series of channels 
\cite{ANNA03,Demenech2006,Rotem2012,Seemann2012}. 
The investigation of the transition between a stable jet and its breakup into a stream of 
droplets is a model paradigm for the much needed control of co-flowing systems,  
ubiquitous in current technological applications.
Regimes for stable jets and unstable dripping jets are 
being studied experimentally,  with theoretical models, and numerical 
simulations \cite{Guillot2007,Utada2007,Couture2011,Lei2011}. 
The breakup of a liquid jet into ever smaller and more 
complex droplets includes the experimental investigation of the effects of 
relative sizes of the channels, as well as channel geometry.  
An attractive experimental technique is recently addressed for a channel which 
is shallow compared to its 
width and length, emptying into a larger channel. The shallow area forces the 
jet to become a ribbon rather 
than a cylinder, and the ribbon remains stable until it flows into a holding 
tank. In this light, the suppression of 
instabilities in multiphase flow by geometric confinement is studied in 
Ref.~\onlinecite{Humphry09}, where the experimental 
work on decreasing the depth of the channel  and  simplified estimates are 
compared to conclude that when 
the depth is sufficiently shallow, the ribbon is stabilized. This idea is used 
for a single step emulsification  \cite{Priest2006,Malloggi2010}. 

In Ref.~\onlinecite{Malloggi2010}, experimental data  for step emulsification
are compared with a model for 
the size of the drops that emerge at the step where the ribbon flows into a 
deeper tank, where the cylindrical necking takes place.  Although this
 is proving to be one of the 
simplest methods to rapidly  
produce   droplets  with controllable sizes and morphologies \cite{Shui2011,Adams2012},
the optimal operating conditions  are not entirely understood.  

The numerical simulation of 
a ribbon or jet sheathed in another liquid, pressure-driven and 
co-flowing through a shallow channel, is a time-dependent simulation because of
the kinematic free surface condition, and the solution quickly reaches a steady state.  A first step 
toward understanding the 
main features is to take advantage of the smallness of the depth of the channel 
compared with the other 
dimensions. Thus, the original governing equations are reduced to the Hele-Shaw 
equations. The key assumptions are given in 
Sec. \ref{sec:equations}.
Our volume-of-fluid (VoF) formulation uses
the balanced-force height-function (HF)
formulation of Ref.~\onlinecite{AB2008}. 
The accuracy for modeling the 
capillary effects is highlighted in this reference.  
Our implementation  is developed 
for a more general class of Hele-Shaw flows of two immiscible viscous liquids than that considered in this paper,
and is novel for the particular regime  where the  interfacial tension force is dominant. The quad-tree 
adaptive mesh refinement\cite{Popinet03,AB2008,AB2009} is  enforced
in regions where much of the important dynamics takes place. The 
balanced-force HF method has the feature of reaching an equilibrium
solution without spurious solutions 
\cite{Prost,FCDKSW2006,AB2009}. For an overview of methods 
for surface-tension dominated multiphase flows, and recent developments, 
including the  phase-field method and the level-set method, the reader is 
referred to  recent publications \cite{Hou2001,Tryggvason2011,Khatri2011}.
 
In Sec.~\ref{sec:methodology}, we present our numerical methodology.
Benchmark computations are given in Sec.~\ref{sec:benchmark}.
These results form a baseline and a standard for numerical accuracy. This is followed in
Sec.~\ref{sec:results} with numerical simulations for the 
experimental conditions of Ref.~\onlinecite{Malloggi2010}.  This reference  derives a
formula for the size of the  neck at the outlet, as a first step toward understanding the mechanism of 
capillary focusing. 
However, this is not a closed formula, and requires empirical input,  
because certain assumptions were made to arrive at a tractable  model.  
Basically, the model reflects inflow and outflow flux 
balances. We  perform numerical simulations in order  to investigate  
whether  the flowfield satisfies those assumptions. 

\section{Governing equations for a volume-of-fluid formulation}
\label{sec:equations}
%
%
\begin{figure}[t]
\centering
\includegraphics[height=0.5\textwidth]{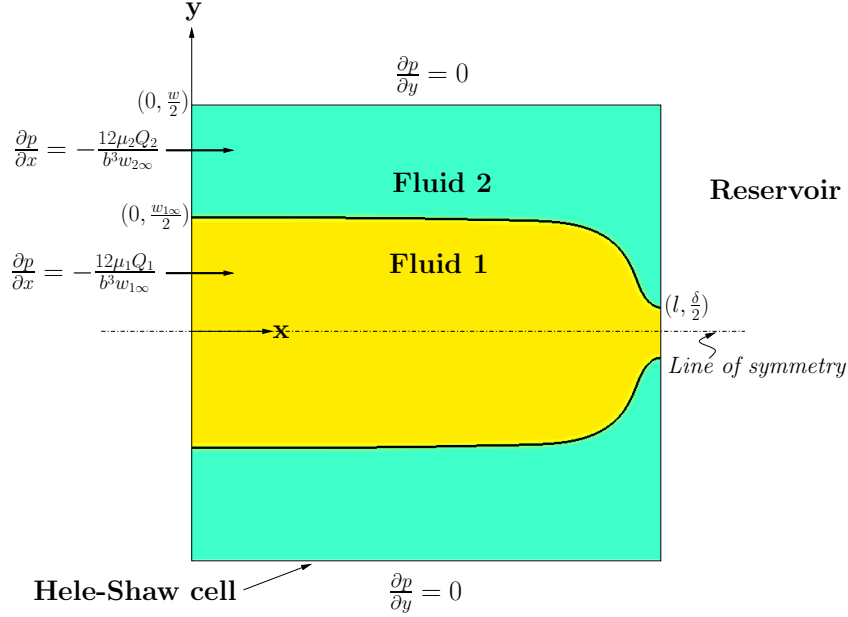}
\caption{Schematic of the flow domain for the Hele-Shaw model.
         The depth is small compared to the width $w$ in the $y$-direction and 
	 the length of the domain $\ell$ in the 
         $x$-direction. At $x=0$, Fluid 1 occupies $|y|\le \frac{w_{1\infty}}{2}$
	 and Fluid 2 occupies  $\frac{w_{1\infty}}{2}\le |y|\le \frac{w}{2}$; $w_{2\infty}=
	 w-w_{1\infty}$. The flow is  
         driven by $\frac{\partial p}{\partial x}$, which is related to 
         the flow rates $Q_i$ for Fluid $i$, $i=1,2 $. The fluids exit at
	 $x=\ell$, where the width $\delta$ of Fluid 1 must be determined as 
         part of the solution.
         The domain is bounded by walls at  $|y| = \frac{w}{2}$.\label{fig:initial}}
\end{figure}
Figure \ref{fig:initial} is a schematic of the flow domain,
$0\le x\le \ell$, $|y|\le \frac{w}{2}$. The depth $b$ in the z-direction is 
small compared with the width $w$ and the length $\ell$. The jet and surrounding
liquid are separated by a wall for $x<0$, and pumped under pressure
through the boundary at $x=0$.  At the exit $x=\ell$,  the width 
of the jet is unknown and is
denoted  by $\delta$. The exit boundary condition is constant pressure
$P_0$, which is a first approximation for the outflow into a reservoir. 
 Capillary effects are expected to decrease  the jet width  across the domain.
The  governing equations prior to a Hele-Shaw approximation are the 3D Stokes equations and incompressibility 
 \begin{equation} {\bf 0}=-\nabla p(x,y,z,t) +\mu \nabla^2{\bf v}(x,y,z,t) +{\bf F}_{ST},\quad
\nabla\cdot{\bf v}(x,y,z,t)=0,\label{eqn:stokes}
\end{equation}
where ${\bf v}=(v_1,v_2,v_3)$, ${\bf F}_{ST}$ denotes
 the body force with the continuum surface tension formulation \cite{Brackbill92}, and
 the  viscosity of  Fluid $i$ is $\mu=\mu_i$, $i=1,2$.
 Fluid 1 occupies $\Omega_1=\{(x,y,z): 0\le x\le \ell, |y|\le h(x,t)< \frac{w}{2}, 0\le
z\le b\}$. Fluid 2 occupies  $\Omega_2=\{(x,y,z): 0\le x\le \ell, h(x,t)\le |y|<
\frac{w}{2}, 0\le z\le b\}$. We denote $\Omega_1\cup\Omega_2=\Omega \subset\mathbb{R}^3$.

The volume-of-fluid formulation identifies each fluid by assigning a VoF function,
\begin{eqnarray}
  \tilde f(x,y,z,t) & = & \left\{ \begin{array}{ll}
        1 & \mbox{in Fluid 1} \\
        0 & \mbox{in Fluid 2}. \end{array} \right.
\end{eqnarray}
 The interface  is calculated by reconstructing the
 curve where the step discontinuity  takes place.  

The sign convention of ${\bf F}_{ST}$ stems from our equilibrium state for
(\ref{eqn:stokes}) where  Fluid 1 (the jet)  bulges into Fluid
2.  Since the pressure in Fluid 1 is higher than in Fluid 2,  $\nabla p$ points into Fluid 1.   
The unit normal ${\bf n}=\nabla \tilde f/|\nabla \tilde f|$  also points into Fluid 1. Therefore,
\begin{eqnarray}
{\bf F}_{ST}=\gamma\kappa \delta_S {\bf n},\label{eqn:fst}
\end{eqnarray}
where    $\delta_S (x,y,z)$ $=|\nabla \tilde f(x,y,z,t)|$ at the 
interface $S$ in the distribution sense \cite{RR1993}.  The curvature is
\begin{equation}
\kappa=-\nabla\cdot {\bf n},\label{eqn:kappa_def}
\end{equation}
where $\nabla\cdot {\bf n}<0$ if the interface bulges 
 into the direction $-{\bf n}$ (into Fluid 2), and $>0$ otherwise.
The fluids  are advected by the velocity field;
\begin{equation}
  \frac{\partial \tilde f(x,y,z,t)}{\partial t} + ({\bf v}\cdot\nabla)\tilde f(x,y,z,t)
= 0.  \label{eq:vofadv_full}
\end{equation}

\subsection{The 2D Hele-Shaw approximation}
Although the 2D Hele-Shaw equations are well known \cite{Ockendon95}, 
we remind the reader of the key ideas in the context of a two-fluid flow.
\begin{enumerate}
\item In the momentum equation for the $x$-$y$ plane, $\rho\frac{D{\bf v}}{Dt}$ is assumed to be negligible compared with $\nabla p$
and $\nabla^2{\bf v}$. The assumption is that $b$ is small, so that $\nabla^2\sim\frac{\partial^2}{\partial
z^2}=O(\frac{1}{b^2})$. This means $p=O(\frac{1}{b^2})$. Thus, $\rho\frac{D{\bf v}}{Dt}$ in the $x$-$y$ plane is assumed to be
smaller order than $O(\frac{1}{b^2})$.

\item The vertical depth between the walls, $b$, is assumed small compared with  the length of the
walls in the $x$-direction,  $\ell$, and the width $w$  in the $y$-direction:  $\frac{b}{w}\ll 1,
\frac{b}{\ell} \ll 1$. The components of the
velocity have magnitudes  $v_1=O(1)$,  $v_2=O(1)$, $v_3=O(b)$.
We define an in-plane depth-averaged  velocity field  ${\bf V}=(V_1(x,y,t),V_2(x,y,t))$,
\begin{equation}
 {\mathbf V}(x,y,t)=\frac{1}{b}\int_0^b (v_1(x,y,z,t), v_2(x,y,z,t)) dz.\label{eqn:capv}
\end{equation}
\item The out-of-plane interface shape is assumed to be semi-circular, with contact angle 180$^o$
at the walls, and radius $b/2$. Thus, the out-of-plane curvature is $2/b$ and contributes $-\gamma
\frac{2}{b}\nabla\tilde{f}$  to the surface tension force. From here, we replace $\kappa$ in (\ref{eqn:fst}) by
$\frac{2}{b}+\kappa(x,y,t)$;
\begin{eqnarray}
{\bf F}_{ST}
= \gamma\left(\frac{2}{b}+\kappa(x,y,t)\right)\nabla\tilde{f}.\label{eqn:newS}
\end{eqnarray}

\item We integrate (\ref{eq:vofadv_full}) with respect to $z$. We have $({\bf v}\cdot \nabla ) \tilde
f(x,y,z,t)=\nabla\cdot({\bf v}\tilde f(x,y,z,t)) $ since $\nabla\cdot{\bf v}=0$. Thus,
\begin{equation}
\int_0^b \left[  \frac{\partial \tilde f(x,y,z,t)}{\partial t} + \nabla\cdot \left( {\bf
v}\tilde{f}(x,y,z,t) \right) \right] dz = 0.\label{eq:vofadv1}
\end{equation}
The first term is $\frac{\partial}{\partial t}\int_0^b \tilde{f}(x,y,z,t)dz$.
The second term is
$\frac{\partial}{\partial x}\left[ \int_0^b v_1\tilde{f}(x,y,z,t) dz\right] $ $+
\frac{\partial}{\partial y}\int_0^b \left[
v_2\tilde{f}(x,y,z,t)  dz\right]  + \int_0^b \frac{\partial}{\partial
z}\left[ v_3\tilde{f}(x,y,z,t)\right] dz$.
We define a depth-averaged VoF function $f(x,y,t)$,
\begin{equation}
f(x,y,t)=\frac{1}{b}\int_0^b \tilde{f}(x,y,z,t)dz.\label{eqn:color_plane}
\end{equation}
The first integral becomes $b\frac{\partial}{\partial t}{f}(x,y,t)$.
The last integral vanishes because $\tilde f$ is bounded, and $v_3(x,y,0,t)=v_3(x,y,b,t)=0$ due to
zero penetration at the walls.  The interface occupies approximately a 
tubular volume with  length  $O(1)$ in the $x$-$y$ plane
 and cross-sectional area of  $O(b^2)$, so that the volume is $O(b^2)$.  The projection in the 
$x$-$y$ plane has area $O(b)$, which shrinks to 0 as $b\to 0$.  We replace $\int_0^b
\left[v_1\tilde{f}(x,y,z,t)\right]dz$ with
${f}(x,y,t)\left[ \int_0^b v_1 dz\right] $, and we define an error $E(x,y,t)$ in $L_\infty$ by
\begin{eqnarray}
E(x,y,t)=\frac{1}{b}\left|\int_0^b v_1\left(\tilde{f}(x,y,z,t)-{f}(x,y,t)\right)  dz\right|.\label{eqn:error}
\end{eqnarray}
We see from  (\ref{eqn:color_plane}) that  $\tilde{f}-f$ is
bounded in the interfacial region, and vanishes away from it.
 The $L_1$ norm of this error is
$\int_{\Omega}E(x,y,t)dx dy\sim b$, which goes to 0 as 
$ b\to 0$.  Therefore, we can approximate  (\ref{eq:vofadv1}) by
$\frac{\partial}{\partial t}{f(x,y,t)}+\frac{1}{b} \frac{\partial}{\partial x} \left[{f}(x,y,t)
\int_0^b v_1 dz\right] +\frac{1}{b} \frac{\partial}{\partial y}\left[{f}(x,y,t) \int_0^b
v_2 dz\right] =0$ in the $L^1$ norm.
In terms of the depth-averaged velocity,
\begin{eqnarray}
\frac{\partial}{\partial t}{f}(x,y,t)+ \frac{\partial}{\partial x}
\left[{f}(x,y,t)V_1 (x,y,t) \right] + \frac{\partial}{\partial y}\left[
{f}(x,y,t)V_2(x,y,t) \right] =0. \label{eqn:adv3}
\end{eqnarray}
Integration of the  incompressibility condition, $\int_0^b (\nabla\cdot {\bf v}) dz=0$, yields
$\nabla_{plane}\cdot{\bf V}=0$,
where $\nabla_{plane}\equiv (\frac{d}{dx},\frac{d}{dy})$.
Therefore, $\nabla_{plane}(f{\bf V})=({\bf V}\cdot\nabla_{plane})f$, which means the advection equation
 (\ref{eqn:adv3}) becomes
\begin{eqnarray}
\frac{\partial}{\partial t}{f}(x,y,t)+({\bf V}\cdot\nabla_{plane}){f}(x,y,t)=0.\label{eqn:adv2}
\end{eqnarray}
Note that  ${\bf V}$, as defined in (\ref{eqn:poiseuille1}),
depends on the curvature, which involves the second derivatives of $\tilde f$. Therefore, (\ref{eqn:adv2}) is
not linear in $f$, and the Courant–Friedrichs–Lewy (CFL) stability condition does not guarantee stability. The stability condition is
complicated by the estimates for ${\bf V}$, which require estimates on the singular contributions of $p$ and
$\nabla\tilde{f}$ at the interface (see Sec.~\ref{sec:stability}).

\item We return to (\ref{eqn:stokes}), and define, for convenience, 
$p^*(x,y,z,t)=p(x,y,z,t)+  \frac{2\gamma}{b}\tilde{f}(x,y,z,t)$.
The classical Hele-Shaw approximation is $v_3=O(b)$, $\frac{\partial}{\partial 
z}=O(\frac{1}{b})$,
 $\nabla^2\sim \frac{\partial^2}{\partial z^2}=O(\frac{1}{b^2})$, as
 $b\to 0$.  The z-component of (\ref{eqn:stokes})  is
$\frac{\partial p^* }{\partial z}=\mu
\frac{\partial^2 v_3}{\partial z^2}-\gamma\kappa(x,y,t)\frac{\partial\tilde f}{\partial z}$.
We assume that the coefficients, 
$\mu, \gamma, \kappa$, are of $O(1)$. We see that  $\frac{\partial p^*}{\partial
z}$ dominates over the other terms if   $p^*=O(\frac{1}{b})$.  With $\frac{\partial p^*}{\partial z}\sim 0$, we conclude that
 $p^*$ is independent of z. Upon consideration of the rest of  (\ref{eqn:stokes}),
$\frac{\partial p^*(x,y,t)}{\partial x}\sim \mu\frac{\partial^2 v_1}{\partial z^2}-\gamma\kappa(x,y,t)
\frac{\partial\tilde f}{\partial x}$, and $\frac{\partial p^*(x,y,t)}{\partial y}\sim
\mu\frac{\partial^2 v_2}{\partial z^2}-\gamma\kappa(x,y,t)
\frac{\partial\tilde f}{\partial y}$, we obtain
  $\frac{\partial p^*(x,y,t)}{\partial x} +\gamma\kappa(x,y,t)
\frac{\partial\tilde f}{\partial x}=O(\frac{1}{b^2})$. Hence,   the z-dependence
disappears and we have  
\begin{equation}
p^*(x,y,t)=p(x,y,t)+  \frac{2\gamma}{b}{f}(x,y,t).\label{eqn:redefinedp}
\end{equation}

Away from the
interface, $\tilde f$ is a constant, and (\ref{eqn:stokes})  reduces to the classical Hele-Shaw equation.
The  $\nabla_{plane}p^*$ terms and
$\nabla\tilde f$ terms  drive the Poiseuille
flow. Also, we find that $p^*=O(\frac{1}{b^2})$.
Together with $v_1=v_2=v_3=0$ at $z=0,b$, we find
$v_1=\frac{1}{2\mu}\left( \frac{\partial p^*(x,y,t)}{\partial x} +\gamma\kappa(x,y,t)
\frac{\partial f}{\partial x}\right)(z^2-bz)$,
$ v_2= \frac{1}{2\mu}\left( \frac{\partial p^*(x,y,t)}{\partial y} +\gamma\kappa(x,y,t)
\frac{\partial f}{\partial y}\right)(z^2-bz)$, and
$v_3=0$. 
The depth-averaged velocities are
\begin{eqnarray}
V_1(x,y,t)=-\frac{b^2}{12\mu} \left( \frac{\partial p^*(x,y,t)}{\partial x} +\gamma\kappa(x,y,t)
\frac{\partial f}{\partial x}\right),\nonumber\\
V_2(x,y,t)=-\frac{b^2}{12\mu} \left( \frac{\partial p^*(x,y,t)}{\partial y}
+\gamma\kappa(x,y,t)\frac{\partial f}{\partial y}\right).\label{eqn:poiseuille1}
\end{eqnarray}
In vector form,   the Hele-Shaw equations are
\begin{equation}
\frac{12 \mu}{b^2} {\ensuremath{\mathbf{V}}} = - \nabla p^* + {\ensuremath{\mathbf{F}}}_{ST},\   0\le x\le \ell,\  |y|\le \frac{w}{2}, \ 
t\ge 0. \label{eq:HS}
\end{equation}

\item In the interface region, the flow does not satisfy the assumption that
$\nabla_{plane}p^*+\gamma\kappa\nabla_{plane}\tilde f$ is a constant with respect to z. However, even though
 (\ref{eqn:poiseuille1}) does not hold pointwise near the interface,  the Hele-Shaw limit is correctly obtained  in the sense of 
distributions (for details, see Ref.~\onlinecite{RR1993}). This  implicitly enforces
the normal stress balance at the interface, which is the continuity of
\begin{eqnarray}
p^*+\gamma\kappa {f}. \label{eqn:continuous}
\end{eqnarray}
  If this is violated, then  the velocity normal to the interface contains a Delta 
function, which contradicts incompressibility. 
\end{enumerate}

\section{Numerical methodology}
\label{sec:methodology}
We implement an iterative procedure toward a unique solution, detailed in this section. In brief, the initial 
interface position determines  the pressure.
With the pressure and interface position known, the velocity is found from (\ref{eqn:poiseuille1}). The
velocity field advects the interface to a new position, and the process repeats
until a steady-state solution is obtained.
The basis for our in-house numerical model is an early version of
Gerris code \cite{Popinet03}.

\subsection{Finite volume discretization}
The computational domain (2D) is initially discretized into square cells with uniform width $\Delta$, 
aligned to the $x$-$y$ coordinates. During the course of a computation, a quadtree adaptive mesh method 
\cite{Khokhlov98} halves $\Delta$  repeatedly  in certain parts of the domain. 
The criteria for  adaptive mesh refinement are based on the pressure gradient, as
well as the location of the interface for the adaptively refined 
solutions. The procedure for the spatial mesh refinement is detailed in Ref.~\onlinecite{Popinet03} and is not repeated here. 

The equation for $p^*$ is formulated from  (\ref{eq:HS}), using $\nabla\cdot {\bf V}=0$,
\begin{equation}
\nabla \cdot \left(\frac{b^2}{12 \mu} \nabla p^* (x,y) \right) =
\nabla \cdot \left(\frac{b^2}{12 \mu} {\ensuremath{\mathbf{F}}}_{ST}\right).
\label{eq:eqpoisson}
\end{equation}
The  weak formulation over cell $(i,j)$
of volume $\Omega_{i,j}$ and bounding surface $S_{i,j}$ is
\begin{equation}
\int_{S_{i,j}} \frac{b^2}{12 \mu} \nabla p^* \cdot \hat{\bf n} \,dS
= \int_{\Omega_{i,j}} \nabla \cdot (\frac{b^2}{12 \mu} {\ensuremath{\mathbf{F}}}_{ST})\,d\Omega,
\label{eq:cvpoisson}
\end{equation}
where $\hat{\bf n}$ is the outward unit normal of $S_{i,j}$.
The finite volume method for the simplest case of uniform grid size $\Delta$ 
yields 
\begin{equation}
\sum_{m} \frac{b^2}{12 \mu_m} \hat{\bf m} \cdot \nabla p^* \Delta^2= \mathcal{D} \Delta^3, 
\end{equation}
 for each cell.  The summation over  $m$ consists of
the four  cell faces, and $\hat{\bf m}$ denotes the outward normal at a face.
$\mathcal{D}$ is the non-zero finite-volume divergence of the
vector field $\frac{b^2}{12 \mu} {\ensuremath{\mathbf{F}}}_{ST}$ defined as
\begin{equation}
\mathcal{D} = \sum_{m} \frac{b^2}{12 \mu_m} \frac{{\ensuremath
F}^m_{ST}}{\Delta},
\end{equation}
where  ${\ensuremath F}^m_{ST}$ is the component of the
surface tension force at the center of the face in the direction of its normal
$\hat{\bf m}$. The computation of $\mu_m$ for interface cells is discussed 
in Sec.~\ref{sec:velocity_cf}.

\subsection{Calculation of curvature}
\label{sec:curvature}
 Within the VoF-based  sharp surface tension
representation,
$\delta_S \hat{\bf n}$ in (\ref{eqn:fst}) is equivalent to $\nabla f$
\begin{equation}
{\ensuremath{\mathbf{F}}}_{ST} = \gamma \kappa \nabla f.
\label{eq:st}
\end{equation}
The curvature is computed at cell centers with the second-order HF method
described in detail in Refs.~\onlinecite{AB2008,AB2009}, and is not repeated here.  
This is currently one of the most accurate 
techniques \cite{Francois2010,Bornia2011}, and contributes to  reduce the overall computational cost.  
At the cell face, the curvature  is interpolated from cell-center values.

\subsection{Boundary conditions}
\begin{description}
\item[Solid wall]
At a solid wall, the boundary condition for the pressure is 
a second-order
discretization of $\nabla p\cdot \hat{\bf n}_{solid}=0$,
i.e.~${\ensuremath{\mathbf{V}}} \cdot \hat{\bf n}_{solid}=0$, where $\hat{\bf n}_{solid}$ is
the unit normal vector to the solid wall. 
The boundary condition for the volume fraction
function at the top and bottom walls is that $f=0$.
\item[Inflow]
With respect to our application in Sec.~\ref{sec:results},  the two fluids are separated by 
a wall up to inflow, so that the inflow boundary condition is $(U_{i\infty},0)$ for Fluid $i$, 
where $i=1,2$.
The parallel flow at inflow is equivalent to prescribed pressure gradients
for both fluids,
\begin{equation}
\frac{\partial p_i}{\partial x} = -
\frac{12 \mu_i Q_i}{b^3 w_{i\infty}}, \quad i=1,2,\label{eqn:qi}
\end{equation}
where subscripts  refer to Fluid $i$,
$w_{i\infty}$ is the width occupied by Fluid $i$ at the inlet,
and $Q_i$ is the inflow rate. 
The boundary condition for $f$ at the inlet is that it is $1$ for
$|y|\le w_{1\infty}$ and 0 otherwise.
\item[Outflow]
At outflow, the pressure is set equal to a reference pressure in the
tank adjoining the Hele-Shaw cell: $p=0$. The boundary condition for $f$ is that
the interface has zero slope: $\nabla f\cdot {\bf n}=\frac{\partial f}{\partial x}=0$.
\end{description}

\subsection{Pressure calculation}
A multigrid V-cycle Poisson solver, accelerated with point relaxation (using 
Jacobi iterations), is used to compute the solution of the system of 
equations generated from (\ref{eq:eqpoisson}).
The adaptive multilevel solver  is described in detail in Ref.~\onlinecite{Popinet03};
in particular, (\ref{eq:eqpoisson}) is solved on a multilevel basis, in which
boundary conditions are interpolated from a previous coarser level solution to
capture the boundary conditions across the multigrid hierarchy.
The criterion for terminating the iterative solution procedure
is that the maximum of the relative residual be smaller than
a specified threshold which is set equal to $10^{-6}$ here. 
The Jacobi pre-smoother with six relaxations per level 
is used. It is known that the convergence of the multigrid method is independent
of the grid size. It is also known that the standard multigrid convergence can 
be degraded in the case of elliptic equations with discontinuous coefficients 
and/or source terms (the condition number of the discretization matrix for
(\ref{eq:eqpoisson})
increases as the ratio of the discontinuous coefficients grows). Since we do not
encounter large viscosity ratios, this degradation does not arise in our
application.

\subsection{Velocity at a cell face}
\label{sec:velocity_cf}
Consider a small discretized cell with volume $v_{cell}$ which is cut by the
interface into a
portion $v_{cell_1}$ occupied by Fluid 1 and $v_{cell_2}$ occupied by Fluid 2. In the cell,
 (\ref{eq:HS}) is satisfied,  and $\mu$ 
is discontinuous at the interface.  In the full Navier-Stokes equations, the velocity is assumed
to be mostly tangential to the interface, and the Hele-Shaw approximation picks up the dominant terms
in the governing equations for this case; for instance, at inflow, this is true, and the in-plane curvature $\kappa$ is small. 
The  regions where this approximation breaks down are  small areas such 
as near the exit, which do not propagate into the bulk of the flow and we check this {\it a posteriori}.  

By projecting  (\ref{eq:HS}) in the direction  normal to the interface, we see that $\frac{\partial p^*}{\partial n}$ is small and $p^*$ is a 
constant in 
the cell. In the direction tangent to the interface,  $p^*$ is continuous, and so is $\nabla p^*\cdot {\bf t}$ where ${\bf t}$ 
denotes a tangent vector to the interface. Therefore, the left hand side of (\ref{eq:HS})  contains  $\mu$ and
 ${\ensuremath{\mathbf{V}}}$ which are both discontinuous, and  the right hand side contains the continuous $p^*$.   We formulate this 
balance by first dividing  by $\mu$, so that both sides have the same singularities. Since  ${v_{cell}}$ is small, and $\nabla p^*$ is 
continuous, the right hand side  is approximated by the 
linearization and we obtain 
\begin{eqnarray}
\int\int_{v_{cell}} {\frac{12}{b^2}\ensuremath{\mathbf{V}}}dxdy=(-\nabla 
p^*+{\bf F}_{ST})\int\int_{v_{cell}}\frac{1}{\mu} dx dy.
\end{eqnarray}
Let us isolate the integral term on the right hand side
\begin{eqnarray}
\int\int_{v_{cell}}\frac{1}{\mu} dx dy=\int\int_{v_{cell_1}}\frac{1}{\mu_1} dx dy+\int\int_{v_{cell_2}}\frac{1}{\mu_2} dx 
dy \nonumber \\
=\frac{1}{\mu_1}v_{cell_1}+\frac{1}{\mu_2}v_{cell_2}=\left(\frac{f}{\mu_1}+\frac{1-f}{\mu_2}\right)v_{cell}.\label{eqn:harmonic1}
\end{eqnarray}
Therefore, (\ref{eq:HS}) gives the average of the velocity over the volume 
\begin{eqnarray}
\frac{1}{v_{cell}}\int\int_{v_{cell}} {\ensuremath{\mathbf{V}}}dxdy=\frac{b^2}{12}(-\nabla 
p^*+{\bf F}_{ST})(\frac{f}{\mu_1}+\frac{1-f}{\mu_2}).\label{eqn:averagev}
\end{eqnarray}
The last bracketed term shows that the viscosity for a mixed cell with index ${i,j}$  is 
calculated from 
the weighted harmonic average 
\begin{equation}
\frac{1}{\mu_{i,j}} = \frac{(1-f_{i,j})}{\mu_2} + \frac{f_{i,j}}{\mu_1}. \label{eqn:harmonic2}
\end{equation} 
Thus, the velocity at the center of a cell face is denoted 
\begin{equation}
  \hat{{\ensuremath{\mathbf{V}}}} = \left\{\frac{b^2}{12\mu}\left(-\nabla p^* +
  {\ensuremath{\mathbf{F}}}_{ST}\right)\right\}_{fc},
 \label{eq:v}
\end{equation}
where the subscript `$fc$' denotes the face-centered quantities.

The harmonic mean  of the viscosities of adjacent cells, say at $(i,j)$ and $(i+1,j)$, are interpolated  to  
compute the viscosity at the  cell face   $(i+1/2,j)$ 
\begin{equation}
\frac{1}{\mu_{i+1/2,j}} = \frac{1}{2}(\frac{1}{\mu_{i,j}} +
\frac{1}{\mu_{i+1,j}}). \label{eqn:harmonic3}
\end{equation}
This viscosity calculation gives a computed nodal velocity that  is closer to the true average 
(\ref{eqn:averagev}) than a 
simple average of the viscosities. This property is demonstrated for the benchmark computation in Sec.~\ref{sec:benchmark1}.

\subsection{Advection of the VoF function}
 The nonlinear advection equation (\ref{eqn:adv2}) presents a  challenge in terms of spatial and 
temporal discretization. An alternative expression is used;
\begin{equation}
  \frac{\partial f}{\partial t} + \nabla_{plane} \cdot ({\ensuremath{\mathbf{V}}}f) = 0.
  \label{eq:vofadv}
\end{equation}
The  normal component of the face-centered velocity 
$\hat{{\ensuremath{\mathbf{V}}}}$ is 
used to advect the VoF function $f$ by solving (\ref{eq:vofadv}). This defines 
new domains for each fluid, and hence  a new position of the 
interface. A piecewise linear interface calculation
is used for the interface reconstruction \cite{Gueyffier98} 
and the Eulerian implicit-explicit scheme described in
detail in Ref.~\onlinecite{Tryggvason2011} is used for the discretization of 
(\ref{eq:vofadv}).

\subsection{Stability conditions}
\label{sec:stability}
It is well known that the explicit formulation of the surface tension force as a
body force is restricted by numerical stability if the governing 
equations are the Euler equations \cite{Brackbill92,Hou1994,Beale1994}. The 
constraint on the time step  is 
$\Delta t\sim (\Delta x)^{3/2}$, and this ensures that  capillary waves 
are not amplified at the 
interface. The constraint for the viscous Navier-Stokes 
equations is found in Ref.~\onlinecite{Galusinski2008} to be 
\begin{equation}
\Delta t\sim (\frac{c_2\mu}{\gamma}\Delta x \Delta t 
+\frac{c_1\rho}{\gamma}\Delta x^3  )^{1/2},\label{eqn:galusinski}
\end{equation}
 where  $c_i$ are positive constants.

In this section, we  clarify the  time constraint for the 
Hele-Shaw equations because it differs from
the aforementioned  estimates. A trivial 
base solution to the  two-fluid Hele-Shaw problem is that of a flat 
interface with zero velocity field.  Consider the effect of small perturbations 
on the length scale of a grid 
cell, localized at the interface, for instance with compact support.  
The corresponding perturbed solution for the interface position and velocity is 
found from linearizing the governing equations about the base 
solution. The kinematic condition is $\frac{D}{Dt}(y-h(x,t))=0$ where $h$ 
represents the perturbed interface position: $y_t=h_t$ or 
\begin{eqnarray}
v=h_t,\label{eqn:vertical}
\end{eqnarray}
where the vertical velocity is $y_t=v$.  The Young-Laplace equation is 
\begin{eqnarray}
-\gamma h_{xx}=[[p^*]],\label{eqn:young}
\end{eqnarray}
where  $[[\ ]]$ denotes the jump  across the interface.
We perform a normal mode analysis, and seek solutions 
proportional to $e^{i\alpha x }$ where $2\pi/\alpha$ is the wavelength,
resolved  to the length scale $\Delta x$ of the discretized cell.
Consider the simplest case, with matched  viscosities, so that   
the steady-state stress balance is   $[[\nabla p^*\cdot{\mathbf n}]]=0$.
   Let the variable   $y$ be shifted to equal 0 at  the interface; in this 
notation,      $[[\frac{\partial p^*}{\partial y}]]=0$.  

In each fluid, the governing equation for the pressure is the Laplace 
equation. 
The solution which decays away from the interface is
\begin{eqnarray}
p^*=
\begin{cases}
(\alpha^2/2)\exp(i\alpha x)\exp(\alpha y),& \text{if} \quad y<0,\\
-(\alpha^2/2)\exp(i\alpha x)\exp(-\alpha y),& \text{if} \quad y>0,
\end{cases}\label{eqn:stability1}
\end{eqnarray}
where, for the investigation of stability, we focus on large $\alpha$. 
The vertical velocity at the interface is, up to a constant factor, 
\begin{equation}
v= -\frac{b^2}{12\mu}{\frac{\partial p^*}{\partial 
y}}=-\frac{\alpha^3b^2}{24\mu}\exp(i\alpha x).
\label{eqn:stability2}
\end{equation}
This equals $h_t$ by (\ref{eqn:vertical}). Substitution of 
(\ref{eqn:stability1}) into
(\ref{eqn:young}) gives   $-\gamma h_{xx}=\alpha^2 \exp(i\alpha x)$. Hence, 
$h_t
=-\frac{\alpha^3b^2}{24\mu}\exp(i\alpha x) =\frac{\gamma b^2 
\alpha}{24\mu}h_{xx}$. Next,
$h_{xx}=-\alpha^2 h$, which gives 
\begin{equation}
h_t=-\frac{\gamma b^2 }{24\mu}{\alpha^3 h} ,\label{eqn:stability3}
\end{equation}  
up to a constant factor.
Thus, the solution is proportional to
$e^{- \frac{\gamma b^2 }{24\mu}\alpha^3 t}$ which is approximated in a 
first-order Euler scheme with the 
Taylor series $ 1- \frac{\gamma b^2 }{24\mu}\alpha^3 t+\dots$.
For a time step $\Delta t$, this truncation is correct if $\frac{\gamma b^2 
}{24\mu}\alpha^3\Delta t \ll
1$. Otherwise, the explicit scheme is unstable.  
The largest wavenumber $\alpha$ which can be numerically resolved is of order 
$1/\Delta x$; therefore, the stability condition is 
\begin{equation}
\Delta t \ll \frac{24\mu}{\gamma b^2}(\Delta x)^3.\label{eqn:stability4}
\end{equation}
 Both this condition and (\ref{eqn:galusinski}) must be satisfied for 
stability of the explicit scheme for 
the viscous 
time-dependent Hele-Shaw equation; our numerical results meet these criteria. 

\section{Benchmark computations}
\label{sec:benchmark} 
Three benchmark computations are presented. The first clearly shows the need for the 
implementation  of the weighted harmonic mean (\ref{eqn:harmonic2})-(\ref{eqn:harmonic3}) for computing the viscosity in a mixed cell.
The second highlights the accuracy of the implemented balanced-force HF method and the calculation of the curvature. Spatial convergence is 
demonstrated by refining the mesh, and tabulating the errors.
The third concerns the accurate implementation of the advection of the VoF function. The 
stability conditions of Sec.~\ref{sec:stability} are enforced to obtain the simulation results.  

\subsection{Two-phase parallel flow driven by a pressure difference: planar 
interface, zero surface tension}
\label{sec:benchmark1}
\begin{figure}[t]
\begin{center}
\includegraphics[trim=20mm 130mm 20mm 125mm, clip=true, width=1\textwidth]{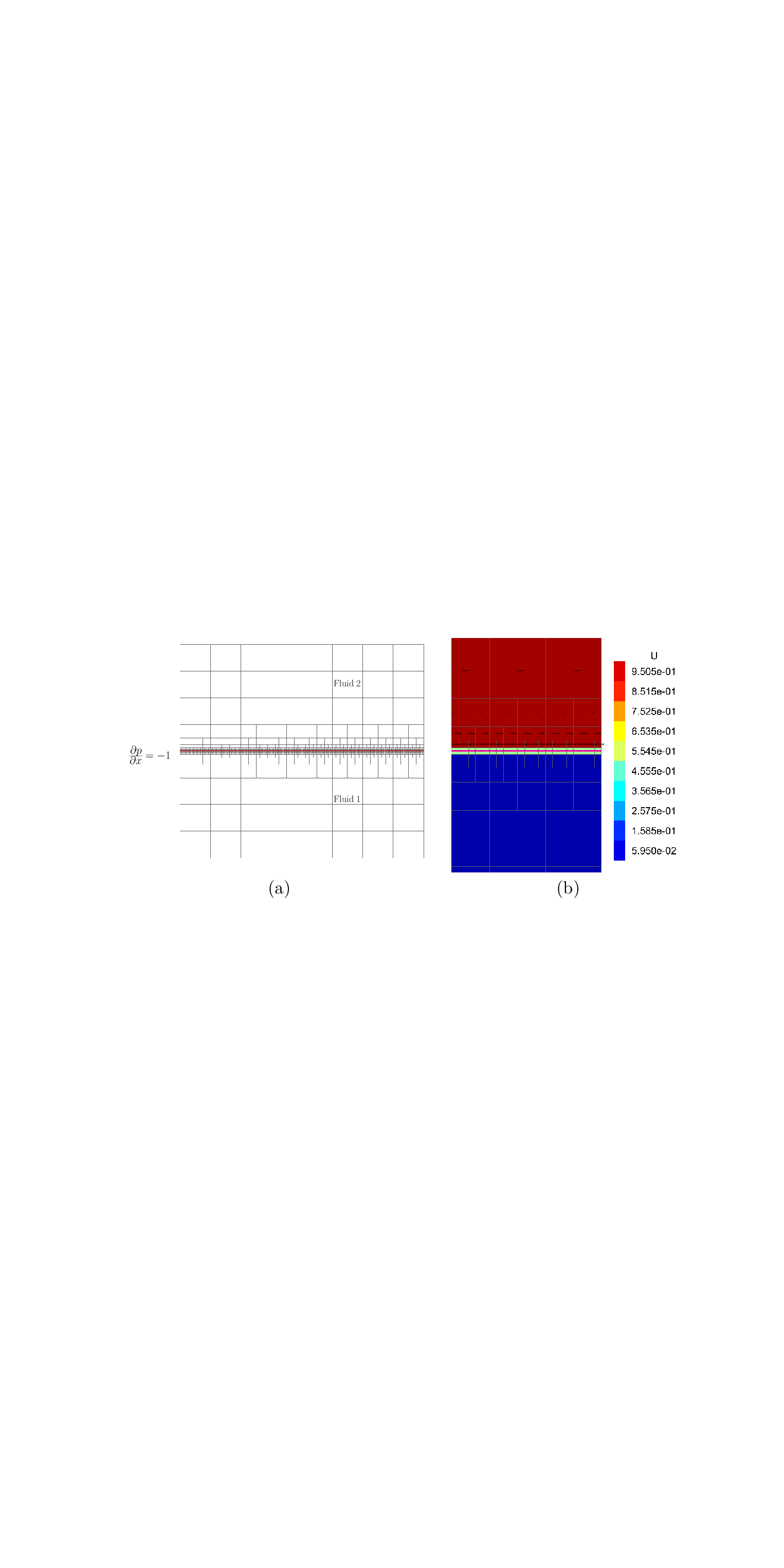}
\end{center}
\caption{(a) Computational domain for the benchmark problem of 
         Sec. \ref{sec:benchmark1} for parallel flow of Fluid 2 over Fluid 1 
	 with a flat interface shown with a solid (red) line.  
	 The domain is 1$\times$1. The boundary 
         conditions are prescribed pressures at inlet and outlet, and zero 
	 vertical pressure gradient at the top and bottom walls of
         computational domain. The interface is adaptively refined, 
         with the smallest mesh size $\Delta=1/256$.
	 (b) Computed velocities in Fluid 1, bottom, and Fluid 2, top; 
         the viscosity ratio $\lambda=100$.}
\label{fig:test-1}
\end{figure}

Consider two fluids of different viscosities in parallel flow. The Hele-Shaw 
equations (\ref{eq:HS})-(\ref{eq:vofadv})
are solved for $\gamma=0$. The boundary conditions are: (i) prescribe pressures 
$p_{\text{in}}$ at the inlet and $p_{\text{out}}$ at 
the outlet such that  a constant pressure difference $\Delta 
p=p_{\text{in}}-p_{\text{out}}$ is maintained;
(ii) zero pressure gradient $\frac{\partial p}{\partial y}=0$  in the direction
normal to the top and bottom boundaries; (iii) prescribe $f$ at the inlet, and 
zero gradient normal to the outlet $\frac{\partial f}{\partial x}=0$, to 
maintain parallel flow at the outlet (see Figure \ref{fig:test-1}(a)).

The exact solution is a planar horizontal interface, with horizontal velocities
\begin{equation}
U_i = \frac{b^2}{12 \mu_i} \frac{\Delta p}{L}.
\end{equation}
The computations are performed for the following values:
$\Delta p/\ell=1$, and the viscosity ratio $\lambda=\mu_1/\mu_2=100$, where
subscripts 1 and 2 refer to the lower and upper fluids, respectively.
Figure \ref{fig:test-1}(b) shows the computed velocities.
The solid (magenta) line shows the location of the interface,
defined to be where the volume fraction of cells cut by the interface is 0.5.
The computed velocities are shown in Figure \ref{fig:test-1}(b), and agree 
with the exact solution in each fluid.
At the interface, the exact  slip velocity is
\begin{equation}
U_1 - U_2 = 
\frac{\mu_2-\mu_1}{\mu_1\mu_2} \frac{b^2}{12} \frac{\Delta p}{L}.
\end{equation}
The computed  slip velocity in cells that are cut by the interface is
\begin{equation}
U_2 + f(U_1 - U_2). 
\end{equation}
Note that the implementation of the weighted harmonic average for the viscosity,
(\ref{eqn:harmonic2}) and (\ref{eqn:harmonic3}), achieves this exact slip velocity.   
On the other hand, if a naive `weighted mean' average of the two viscosities
is used to compute the viscosity of a mixed cell, and a simple
average of cell center viscosities is used to interpolate the 
viscosity to the face of the cell, then the slip velocity
of the mixed cell would be strongly shifted towards the more viscous fluid 
(Fluid 1 in this example). We avoid this inaccuracy.

\subsection{Circular interface in equilibrium: non-zero surface 
tension}
\label{sec:benchmark2}
\begin{figure}[t]
\centering
  \includegraphics[width=0.5\textwidth]{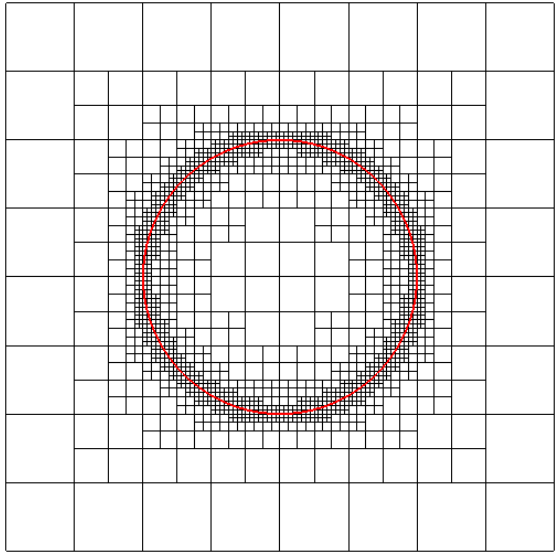}
\caption{Benchmark computation from Sec.~\ref{sec:benchmark2}. Fluid 1 
         occupies the interior of the circle, and Fluid 2 occupies 
         the exterior. The circular interface is shown as a solid (red )
         line with radius 0.25, at the center of a 1$\times$1 
         computational domain. The interface is adaptively refined; here,
         the smallest mesh size $\Delta=1/128$.}
\label{fig:test-2}
\end{figure}
Consider a circular drop placed at the center of a square 
computational domain that is initially at rest,
as shown in Figure \ref{fig:test-2}.
The inflow and outflow boundary conditions are zero 
normal pressure gradients.  The numerical simulation presented here is a test
for the accuracy of the computation of the  interfacial tension force.
The initial configuration is a solution
of (\ref{eq:eqpoisson}) and satisfies the 
Young-Laplace condition, $[[p]]=\gamma \kappa$.

%

\begin{table}[t]
\caption{Convergence results for the benchmark problem for Sec.~\ref{sec:benchmark2}.
         The norms $L_1$, $L_2$, and $L_{\infty}$  of 
         the velocity, and the pressure jump across the interface are shown 
	 as a function of mesh refinement.\label{tab:test-2} }
\begin{center}
\begin{ruledtabular}
\begin{tabular}{ c  c  c  c}
    & $\Delta$=1/32 & $\Delta$=1/64 & $\Delta$=1/128 \\
  \hline
  $L_1$  & 3.416e-06  & 5.686e-07  & 1.326e-07  \\
  $L_2$  & 5.234e-06  & 8.882e-07  &  2.083e-07 \\
  $L_{\infty}$  & 1.680e-05  & 2.879e-06  & 6.461e-07  \\
  $p_{1}-p_{2}$  & 4.03114  & 4.00750  & 4.00179  \\
\end{tabular}
\end{ruledtabular}
\end{center}
\end{table}

The computations  are started at the discretized  equilibrium solution, with 
zero velocity and a circular interface of radius $r=0.25$.
The viscosity ratio is chosen as $\lambda=1$, the interfacial tension is
$\gamma=1$, and the time step is $\Delta t=10^{-6}$.
Table \ref{tab:test-2}  presents the spatial convergence based on the
$L_1$, $L_2$, and $L_{\infty}$
norms of the velocity field, and the pressure jump across the interface at the
5000th time step. ($p_{1}$ is the averaged pressure for cells with $r<0.25$
and  $p_{2}$ is the averaged pressure for cells with $r>0.25$). 
It is clear that  the velocity field decreases to zero with the mesh size 
($\Delta=\frac{1}{32}, \frac{1}{64}, \frac{1}{128}$).  
At the 5000th time step, the computed velocity  is not zero  because  the 
numerically computed interface shape has not reached an equilibrium. 
At each mesh resolution, there is a difference between the exact 
circular shape and the numerically computed interface shape; however, after a 
sufficient number of time steps,  
our balanced-force HF method has the feature of reaching the 
equilibrium velocity of zero to machine precision
\cite{AB2009}. At a fixed time step, the non-zero velocity 
diminishes at a second-order rate with mesh
refinement. Additionally, the pressure jump across the interface approaches the
exact value with second-order accuracy. Hence, Table \ref{tab:test-2} 
demonstrates that our numerical methodology for the interfacial tension force 
yields converged solutions.

\subsection{Translation of a viscous droplet  in an unbounded Hele-Shaw flow}
\label{sec:benchmark3}
\begin{figure}[t]
\centering
  \includegraphics[width=0.75\textwidth]{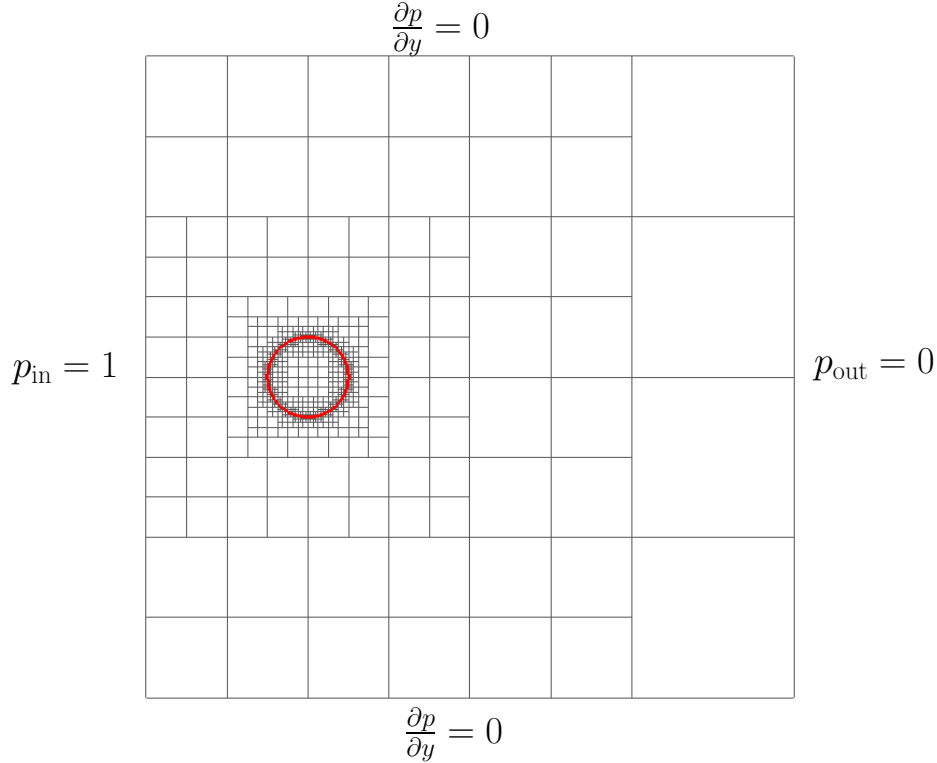}
\caption{A droplet of Fluid 1, with a circular interface of radius 0.0625, is
         placed in a 1$\times$1 computational domain. The interface is shown as
	 a solid (red) line. The domain is adaptively refined while the
	 droplet translates in response to the flow of the surrounding
	 Fluid 2.}
\label{fig:test-3}
\end{figure}

Here we consider translational motion of a highly viscous droplet with high
interfacial tension in an unbounded Hele-Shaw cell with an imposed uniform flow 
far from the droplet. The exact solution is the translation of the droplet. The
fluid within the droplet moves as a rigid body with no recirculation. 
The boundary conditions for the pressure are: (i) at the upper and lower 
boundaries, the pressure gradient  $\partial p / \partial y=0$;  (ii)  The
pressure at the inlet is prescribed by the constant $p_{\text{in}}$ and at the
outlet by the constant $p_{\text{out}}$ so that sufficiently far away from the 
drop, $\frac{\partial p}{\partial x}<0$. 
The boundary conditions for the VoF function are; (i) $f$ is prescribed to be
Fluid 2 at the inlet and top and bottom walls, and $df/dx=0$ at the outlet.
Computationally, the droplet must be much smaller than the cell to guarantee a 
constant pressure gradient far from the droplet.

We check the velocity of droplet translation. In this case it can be shown that
a circle is an exact solution for the steady shape of the translating
droplet of an arbitrary surface tension\cite{Gupta1999,Leshansky}
 with the corresponding pressure distribution given in polar coordinates
($r,\theta$) by
\begin{equation}
p_{\text{drop}}= \left(\frac{2}{b} + \frac{1}{a}\right)\gamma+ 
\frac{2\mu_1}{\mu_1+\mu_2}\frac{-\Delta p}{L}r\cos\theta,
\label{eq:drop-pressure}
\end{equation}
\begin{equation}
p_{\text{matrix}}= \left(1+\frac{a^2}{r^2}\frac{\mu_1-\mu_2}{\mu_1+\mu_2}\right)
\frac{-\Delta p}{L}r\cos\theta,
\end{equation}
where $a$ is the droplet radius, $\theta$ is measured from the direction
of the applied pressure gradient, and $r$ represents the radial distance from the
center of the drop. The steady (rigid body) translational velocity of the
circular drop is
\begin{equation}
U= \frac{b^2}{12\mu_1}\frac{2\mu_1}{\mu_1+\mu_2}\frac{\Delta p}{L}.
\label{eq:trans-vel}
\end{equation}

\begin{figure}[t]
\centering
\includegraphics[width=1\textwidth,trim= 25mm 130mm 0mm 120mm,clip=true]{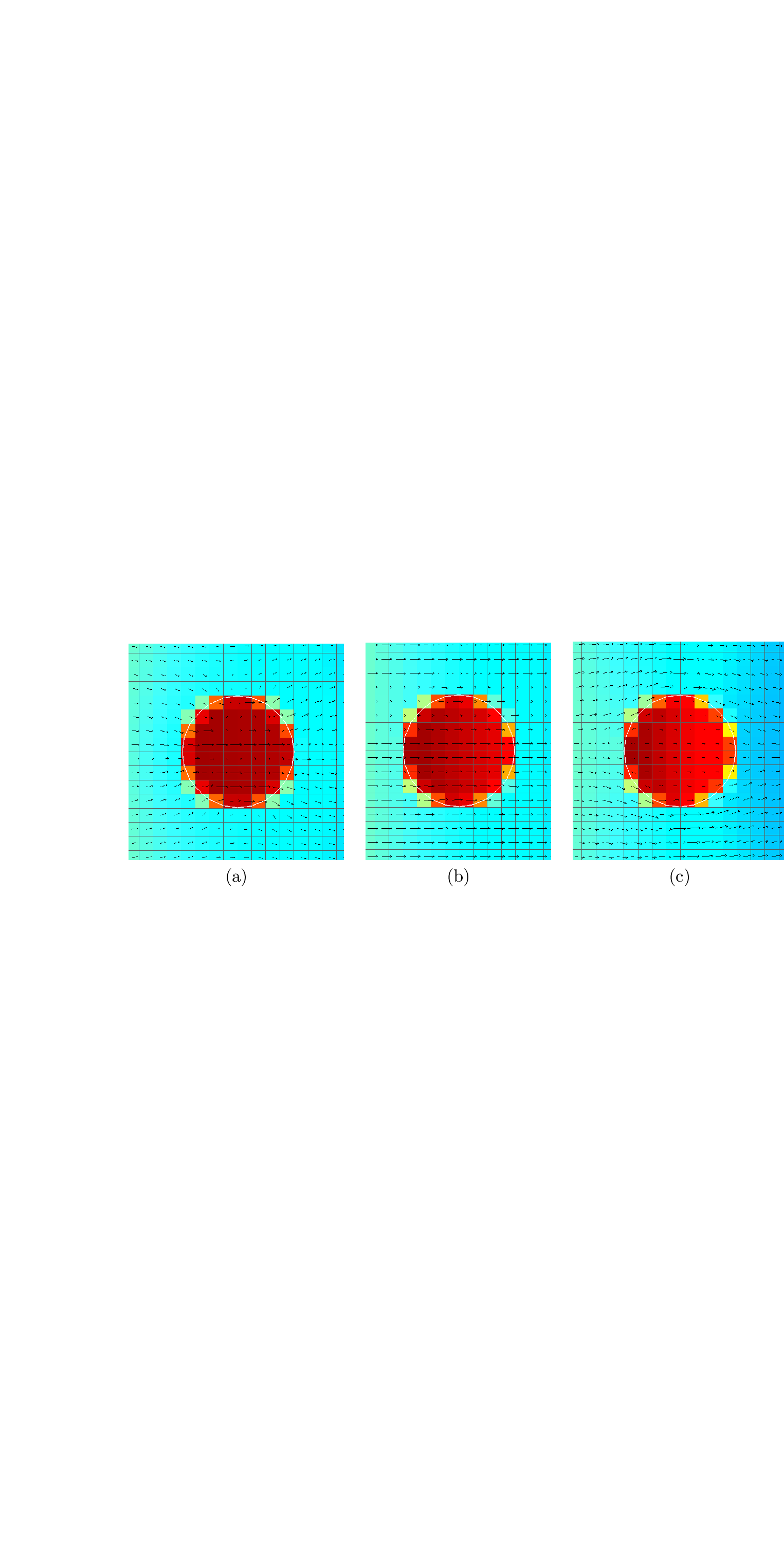}
\caption{Linear pressure distribution and the velocity field
         for $\lambda=0.1$ (a), $1$ (b), and $5$ (c).  
         Contours  show the pressure field distribution
         with a maximum value colored in dark (red) and a minimum value colored
         in light (blue).}
\label{fig:test3-pressure-velocity}
\end{figure}

\begin{table}[t]
 \caption{The steady translational velocity of the circular drop for different
          viscosity ratios compared with the predicted velocity by 
	  (\ref{eq:trans-vel}).\label{tab:trans-vel}} 
\begin{ruledtabular}
\begin{tabular}{ c  c  c  c}
      & $\lambda=0.1$ & $\lambda=1$ & $\lambda=5$ \\
  $U_{\text{computed}}$  & 1.90  & 1.005  &  0.41 \\
  $U_{\text{theory}}$  & 1.82  & 1.0  &  0.33 \\
\end{tabular}
\end{ruledtabular}
\end{table}
Here we consider a drop of radius $0.0625$ placed in a 1$\times$1 computational
domain. We check that the radius of the drop is small enough so that it will not
affect the far field pressure distribution (Figure \ref{fig:test-3}). We set $\gamma=1$, $\Delta p/\ell=1$,
and vary the viscosity ratio $\lambda$ from $0.1$ to $5$. The comparison between the numerically
computed steady translational velocity of the drop and 
 (\ref{eq:trans-vel}) is shown in Table \ref{tab:trans-vel}.
 We observe that the comparison is 
improved  with better resolution
of the flowfield. 
Figure \ref{fig:test3-pressure-velocity} shows the snapshots of the numerical
simulations for viscosity ratios of $0.1$, $1$, and $5$. More detail about the velocity and pressure
fields follow. 

First, it is evident that the initially circular  drop (solid white line) 
remains in equilibrium for all cases.
We also note that the translational velocity (\ref{eq:trans-vel}) does not depend on the
interfacial tension; this is confirmed with numerical simulations at $\gamma=0.1$, $0.01$.
Secondly, Figure \ref{fig:test3-pressure-velocity}  shows the pressure
distribution inside the drop (color contours). 
For $\lambda = 1$, the pressure gradient
inside the drop is simply $\partial p/\partial x = -\Delta p/\ell=-1$,
in agreement with the theoretical value  (\ref{eq:drop-pressure}). This
 is the pressure gradient imposed  at the far field.
According to (\ref{eq:drop-pressure})  for $\lambda=0.1$, the pressure
gradient inside the drop is smaller and for $\lambda=5$, the pressure
gradient inside the drop is larger than the imposed
far field pressure gradient $\Delta p/\ell=1$. Figures \ref{fig:test3-pressure-velocity}(a)-\ref{fig:test3-pressure-velocity}(c)
support this prediction. It is also noted
that the pressure distribution inside the drop for $\lambda=0.1$ is approximately a constant
because it should be close to 
the pressure distribution of an inviscid drop ($\lambda \to 0$)
in a Hele-Shaw cell which is known to have a steady translational 
velocity of 2. 
Thirdly, Figure \ref{fig:test3-pressure-velocity} shows the velocity field
inside and outside the drop. Clearly,
each drop undergoes a rigid body translation with a steady
velocity that is predicted by (\ref{eq:trans-vel}),
i.e.~the velocity inside the drop is a zero velocity field in a frame of 
reference moving with the drop steady-state velocity. 

\section{Pressure driven flow of a co-flowing ribbon}
\label{sec:results}
We turn to the pressure-driven flow of a jet or ribbon of one fluid co-flowing with a second fluid
which is shown in Figure \ref{fig:initial}, through a channel that is much wider than it is deep. 
For this flow, it is possible to make the jet form a tongue at the exit boundary, 
where the width of the tongue is extremely small. The production of small droplets, on the order of the depth of the
channel, ensues in the reservoir, and is known as capillary focusing. Of practical importance is a simple
estimate for the jet width $\delta$ at the exit; Ref.~\onlinecite{Malloggi2010}
is a first attempt to estimate $\delta$  and  compare with the experimental data.  However, capillary 
focusing occurs at small capillary numbers, and the comparison
appears to suffer in this regime, while the comparison for O(1) capillary number is satisfactory. Questions
arise about the  
 assumptions that are built into their estimate. This section clarifies this issue by directly 
interrogating  the flow with numerical simulations.
\subsection{A rough estimate for jet width at exit}
\label{sec:assumptions}
A summary of the main assumptions in the prior estimate\cite{Malloggi2010} follows: 
\begin{enumerate}
\item  The $x$-component of the Hele-Shaw equation without the presence of  the interface \cite{Ockendon95} is
$\frac{12\mu}{b^2}V_1(x)=- \frac{\partial P(x,y)}{\partial x} $. This equation is integrated  
along two streamlines from the inflow to the outflow. Together with the outflow condition $P_1(0)=P_2(0)$, 
the result is 
 $-P_i(-\infty)+P(0)=\frac{12\mu}{b^2}\int_{-\infty}^0  V_1(x,y) dx$.
The two streamlines are (i) along the   centerline  
$y=0$ (Fluid 1), and (ii) at the wall $y=\frac{w}{2}$ (Fluid 2). 
Subtraction of one equation from the other yields
\begin{eqnarray}
-P_2(-\infty)+P_1(-\infty)=\frac{12}{b^2}\int_{-\infty}^0 D(x)  dx   
,\nonumber\\
D(x)=\mu_2U_2(x,\frac{w}{2})-\mu_1 U_1(x,0).\label{eqn:Malloggi2}
\end{eqnarray}
This relates the quantities at outflow  to  the prescribed 
inflow quantities, but  in order to simplify this further, 
a decay property is imposed on $D(x)$. 
\item  At the interface between the fluids, the jump in the normal stress is balanced by surface tension effects. 
Here, because the depth $b$ is small, the in-plane curvature is neglected in comparison with the out-of-plane 
($y$-$z$) value $\frac{2}{b}$, which 
originates from the semi-circular diameter. Hence, at inflow,
\begin{eqnarray}
\frac{2\gamma}{b}=P_{2}(-\infty)-P_1(-\infty).\label{eqn:Malloggi3}
\end{eqnarray}
Substitution  into  (\ref{eqn:Malloggi2}) yields
$\int_{-\infty}^0 D(x) dx =\frac{-\gamma b}{6}$. 
Thus, the left hand side is a convergent improper integral. Therefore, the integrand must 
decay sufficiently fast to 0 at the lower end of the integration. This is taken one step further 
 with the assumption that there is a decay length $\ell_0$, defined by 
$\int_{-\infty}^0 D(x) dx=\ell_0 D(0)$. 
This leads to a tractable expression  
\begin{eqnarray}
\ell_0 D(0)  =\frac{-\gamma b}{6}, \quad 
D(0)=\mu_2 U_2(0)-\mu_1 U_1(0).\label{eqn:Malloggi4}
\end{eqnarray}  
This and (\ref{eqn:Malloggi2}) are equations that link inflow (prescribed) and
outflow (unknown) quantities.
\item  Since the interior flow is not known {\it a priori}, a flux conservation is imposed
\begin{eqnarray}
 U_{1\infty}w_{1\infty}=U_1(0)\delta,  \quad
U_{2\infty}w_{2\infty}=U_2(0)(w-\delta). \label{eqn:fluxes} 
\end{eqnarray}
Since $w_{i\infty}$ denotes the width of Fluid $i$ at inflow, we have   
$w_{1\infty}+w_{2\infty}=w$. 
Combined with (\ref{eqn:Malloggi4}),   the unknown outflow velocities are 
eliminated and we have $D(0)$ in terms of the  inflow data.
Substitution in (\ref{eqn:Malloggi3}) gives the estimate for $\delta$.
The final equation is $z^2\beta-z(1+Ca_M)+Ca_M=0$, where  
\begin{eqnarray}
 \beta=\frac{w_{1\infty}}{w},\quad 
 z=\frac{\delta}{w_{1\infty}}=\frac{\delta}{w\beta},\label{eqn:Malloggi6}
\end{eqnarray}
and 
\begin{eqnarray}
  Ca_{M}=\mu_2 \frac{6 U_{2\infty}\ell_0}{b\gamma}=\mu_1  \frac{6
U_{1\infty}\ell_0}{b\gamma}.\label{eqn:capillary}
\end{eqnarray}
The equivalence of the two formulas follows from $D(-\infty)=0$.
The usefulness of  $Ca_M$ is limited because it  depends on the unknown decay factor $\ell_0$. 
The estimate becomes
\begin{eqnarray}
z=\frac{(1+Ca_M)}{2\beta}\left( 1-\sqrt{1-\frac{4Ca_M 
\beta}{(1+Ca_M)^2}}\right).\label{eqn:Malloggi7a}
\end{eqnarray}
\item Some observations: (i) If the interfacial tension is large enough,  then $ Ca_M \ll 1$, and  (\ref{eqn:Malloggi7a})
predicts that $\delta$ decreases at the rate $z\sim Ca_M$.
(ii) If the interfacial tension is small, then $Ca_M \gg 1$ and  (\ref{eqn:Malloggi7a})
predicts that  $\delta$ is the same as the inflow width: $z\sim 1$. Indeed,  any
reasonable estimate must predict that the
interface becomes flatter through the domain with increasing $Ca_M$.  
(iii) The free parameter $\ell_0$, set to $w/2$, is found to be a good fit to experimental data in Figure 2(c) of Ref.~\onlinecite{Malloggi2010}. 
\end{enumerate}

\subsection{Numerical simulations}
\label{sec:numerical_malloggi}

\begin{figure}[t]
\centering
\includegraphics[width=100mm]{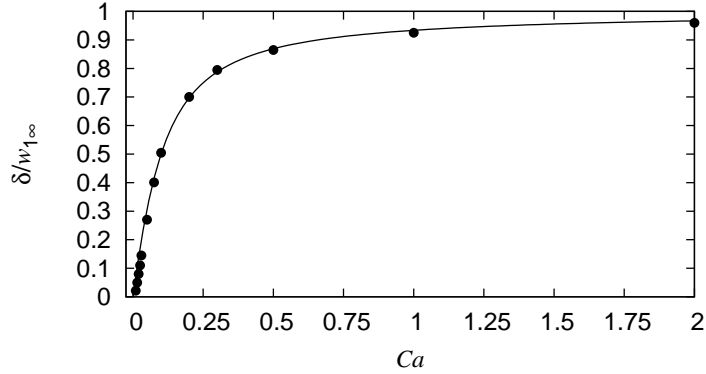}
\caption{\label{fig:Malloggi} The width of the jet $\delta$ upon exit scaled with $w_{1\infty}$ as a
         function of the capillary number $Ca$. $w/b=10$, $k=1$, and
	 $w_{1\infty}/b=5$. Numerical results ($\small{\bullet}$) and
	 an apparent fit (--) using (\ref{eqn:Malloggi7a}) with $\ell_0/w=0.125$.}
\end{figure}
The flow conditions of Ref.~\onlinecite{Malloggi2010} are numerically simulated with
$\mu_1=\mu_2$, and $dp_1/dx=dp_2/dx$ at the inflow.
The exact solution which provides the inflow conditions is $w_{1\infty}/w=1/(1+k)$ where
 $k=(Q_{2\infty}\mu_2)/(Q_{1\infty}\mu_1)$, and the flow rates $Q_{i\infty}$ for $i=1$, $2$ are defined in (\ref{eqn:qi}).
Our capillary number  is defined by
\begin{eqnarray}
 Ca=\frac{\mu_1 U_{1\infty}}{\gamma},
\label{eqn:our_ca}
\end{eqnarray}
and  is not  $Ca_M$ in (\ref{eqn:capillary}). 
The channel aspect ratio $w/b$ is assumed large, and the Hele-Shaw approximation is expected to be more accurate as $w/b$ increases.
Figure \ref{fig:Malloggi} reports the results of the numerical simulations ($\small{\bullet}$)
for $\delta$ as a function of the capillary number $Ca$ for $w/b=10$, 
$k=1$, and $w_{1\infty}/b=5$, together with an apparent fit (--) using (\ref{eqn:Malloggi7a}),
with $\ell_0/w=0.125$.
Note that  the interface remains straight when the interfacial tension is
small; this accounts for  $\delta/w_{1\infty}\approx 1$ when $Ca$ is large. This is also found in Figure 2(c) of Ref.~\onlinecite{Malloggi2010}, where the trend at
$Ca=O(1)$ is used to choose $\ell_0/w$. The problem with this method is that on the scale of the figure,
the results for $Ca\ll 1$ are too small to be discerned. The line in Figure~\ref{fig:Malloggi} 
represents
the estimate (\ref{eqn:Malloggi7a}) with a choice of
$\ell_0/w= 0.125$, and shows an apparent fit. However,  the jet forms a tongue only if  $Ca\ll 1$,
and we next focus on this regime. 

\begin{figure}[t]
\centering
\includegraphics[width=90mm]{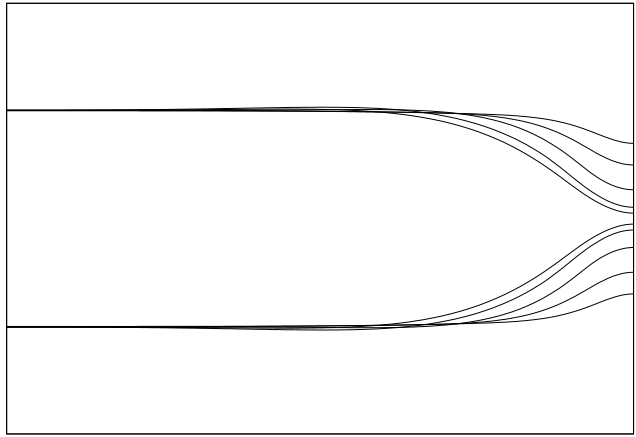}
\caption{Steady-state jet solution for $Ca=0.01$, $0.025$, $0.05$, $0.1$, and $0.2$ 
         (small to large $\delta$), for $w/b=10$, $k=1$, and $w_{1\infty}/b=5$.
	 The curvature at the outlet is  $\kappa=2/b$.}
\label{fig:profiles}
\end{figure}
Figure \ref{fig:profiles} shows the numerically
computed steady-state shapes of the jet 
for $w/b=10$, $k=1$, and $w_{1\infty}/b=5$. The capillary number is varied from $0.01$ to $0.2$
in order to show the same trend as in the available experimental
data: the jet narrows more at the outlet as the capillary number decreases
\cite{Priest2006, Malloggi2010}.

\begin{figure}[t]
\centering
\includegraphics[width=100mm]{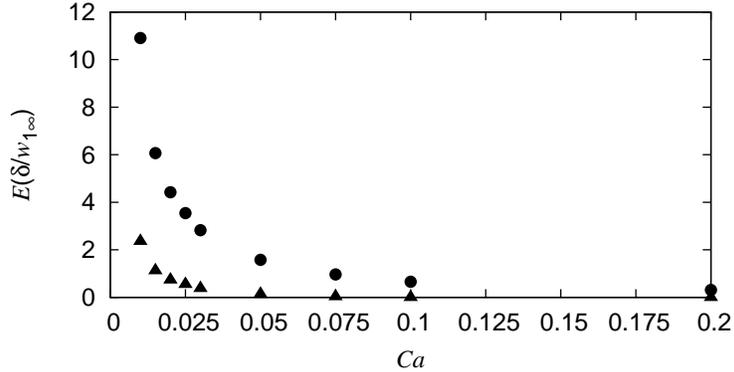}
\caption{Relative error $E(z)$ of (\ref{eqn:Malloggi7a}) as a function of $Ca$
         for $\ell_0/w=0.5$ ($\bullet$) and $0.125$ ($\small {\blacktriangle}$).}
\label{fig:error-ca}
\end{figure}
It is  informative to present 
 the relative error of the model $z_{m}$ 
(\ref{eqn:Malloggi7a}) with respect to the exact computed values $z_{c}$, defined by
$$E(z)=\frac{z_c-z_m}{z_c}.$$
Figure \ref{fig:error-ca} shows $E(z)$
as a function of $Ca$. If $Ca$ increases past $0.1$,  then the interfacial tension is weak and the interface  tends to
stay undeformed;  
$\delta/w_{1\infty}$ naturally approaches 1, no matter the choice of $\ell_0/w=0.5$ ($\bullet$) and
0.125 ($\small {\blacktriangle}$).
However, if $Ca$ is small, then Figure \ref{fig:error-ca} shows that 
the relative errors  are  large no matter what value of $\ell_0/w$ is picked. Therefore, the assumption
in Sec.~\ref{sec:assumptions} that the  decay length $\ell_0$ is comparable to the channel width $w$
is not correct. With hindsight, we see that the assumption of such a large
 decay region is incompatible with the assumption in  Sec.~\ref{sec:assumptions} that the velocity is uniform in each fluid along 
$x=$ constant, and also with the assumption that the $y$-component of velocity has no role in the derivation of  
(\ref{eqn:Malloggi7a}).  The numerical simulations also confirm the expectation that
 there is a significant 
$y$-component of velocity in a much smaller `decay region' very close to the exit.  

\begin{figure}[t]
\begin{center}
\includegraphics[trim=40 100mm 0 90mm 0,width=1.05\textwidth]{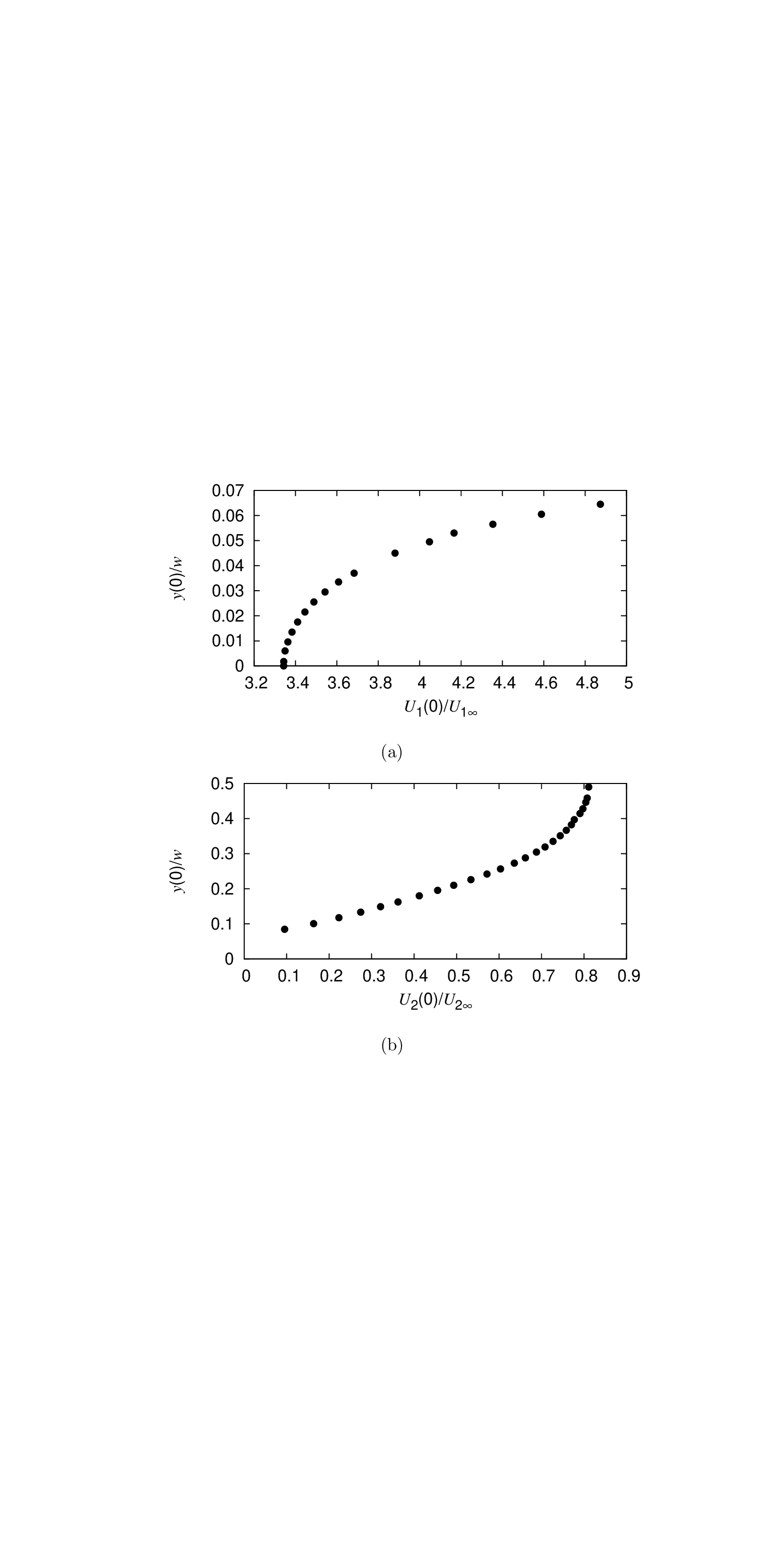}
\end{center}
\caption{(a) $U_1(0)$ and (b) $U_2(0)$, normalized by 
         $U_{1\infty}$ and $U_{2\infty}$, respectively,
	 along the outflow boundary $x=0$. $Ca=0.05$, $w_{1\infty}/b=5$, and $w/b=10$. 
	 $\mu_1/\mu_2=1$. $k=1$.}
\label{fig:v12}
\end{figure}

\begin{figure}[t]
\begin{center}
\includegraphics[scale=0.45,trim=0 0 50mm 0,clip=true]{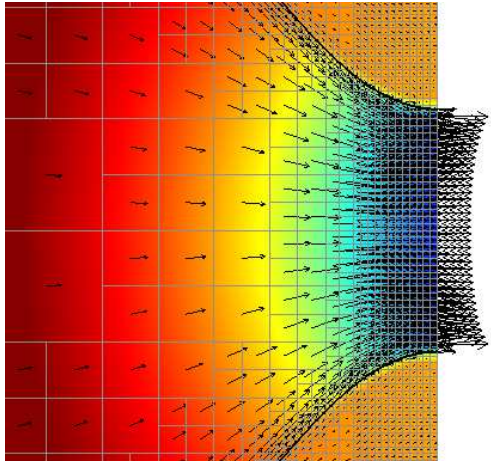}
\end{center}
\caption{Flow field and the pressure
         distribution in the focusing region for
         $Ca=0.05$, $w_{1\infty}/b=5$, and $w/b=10$; 
	 $\mu_1/\mu_2=1$ and $k=1$. The pressure contours 
	 show that at the outflow boundary, the pressures in both phases equilibrate.}
\label{fig:velocity}
\end{figure}

Figure \ref{fig:v12} confirms that the flux does not satisfy  the assumption (\ref{eqn:fluxes}), 
which feeds into (\ref{eqn:Malloggi7a}).  The figure shows numerically computed velocities
$U_1(0)$ and $U_2(0)$ (here we use $x=0$ to denote the outflow position and $x=-\infty$ for inflow), normalized by 
$U_{1\infty}$ and $U_{2\infty}$,
respectively, along the outflow boundary.
The numerical results show a complex non-uniform velocity field at  outflow, which is ignored if only the
inflow and outflow flux conditions are used in the theoretical analysis. 
An interesting feature of the velocity distribution in Figure \ref{fig:v12}
is a strong slip between the inner and outer fluids. There is more
than an order of magnitude difference between the inner phase and outer phase velocity
at the Fluid 1-Fluid 2 interface.  
The corresponding velocity field is shown  in Figure \ref{fig:velocity}.
 The square cells depicted in Fluid 1 illustrate the spatial 
discretization for the adaptive mesh refinement. The shading indicates the pressure
distribution.  
The numerical results of Figures \ref{fig:v12} and \ref{fig:velocity} are for $Ca=0.05$
where the flow focusing is moderate compared with the stronger focusing at smaller values of $Ca$ 
shown in Figure \ref{fig:profiles}. 
On the other hand, if  $Ca$ is larger than 0.1,  we find that the assumption of gentle flow variation
from inflow to outflow, which was used to obtain  (\ref{eqn:Malloggi7a}), is reasonable.

\begin{figure}[t]
\begin{center}
  \includegraphics[trim=30mm 100mm 0 120mm,width=7.in]{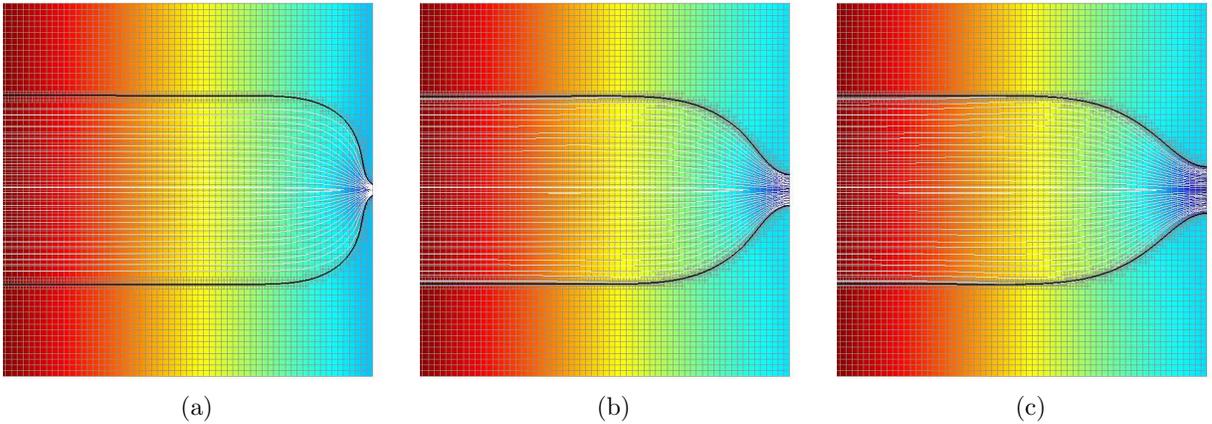}
\end{center}
\caption{Effect of surface tension and channel depth on self-focusing.
         Contours depict the pressure field distribution
         with a maximum value colored in dark (red) and a minimum value colored
         in light (blue).
	 The adaptive mesh and the streamlines in the inner stream
         are also shown. (a) $Ca=0.016$, $w_{1\infty}/b=10$, and $w/b=20$. (b) $Ca=0.033$,
         $w_{1\infty}/b=5$, and $w/b=10$. (c) $Ca=0.052$, $w_{1\infty}/b=3.33$, and 
	 $w/b$=6.66. $\mu_1/\mu_2=1$ and $k=1$.}
\label{fig:numerics1}
\end{figure}

Figure \ref{fig:numerics1}  shows the numerical results of the steady-state shapes of the interface
at three values of the capillary number, and confirms 
that the focusing effect is stronger for smaller capillary numbers. 
The significant 
focusing is evident at $Ca = 0.016$ (Figure \ref{fig:numerics1}(a)); i.e.~high surface tension
yields improved self-focusing. Figure \ref{fig:numerics1} also shows an important feature 
that by decreasing the channel depth, the narrow jet  develops
a sharp tip at the outflow boundary. The pressure distributions are shown in 
Figure \ref{fig:numerics1} for varying $Ca$ and $w/b$.

\begin{figure}[t]
\begin{center}
\includegraphics[trim=43mm 137mm 35mm 130mm,width=0.7\textwidth]{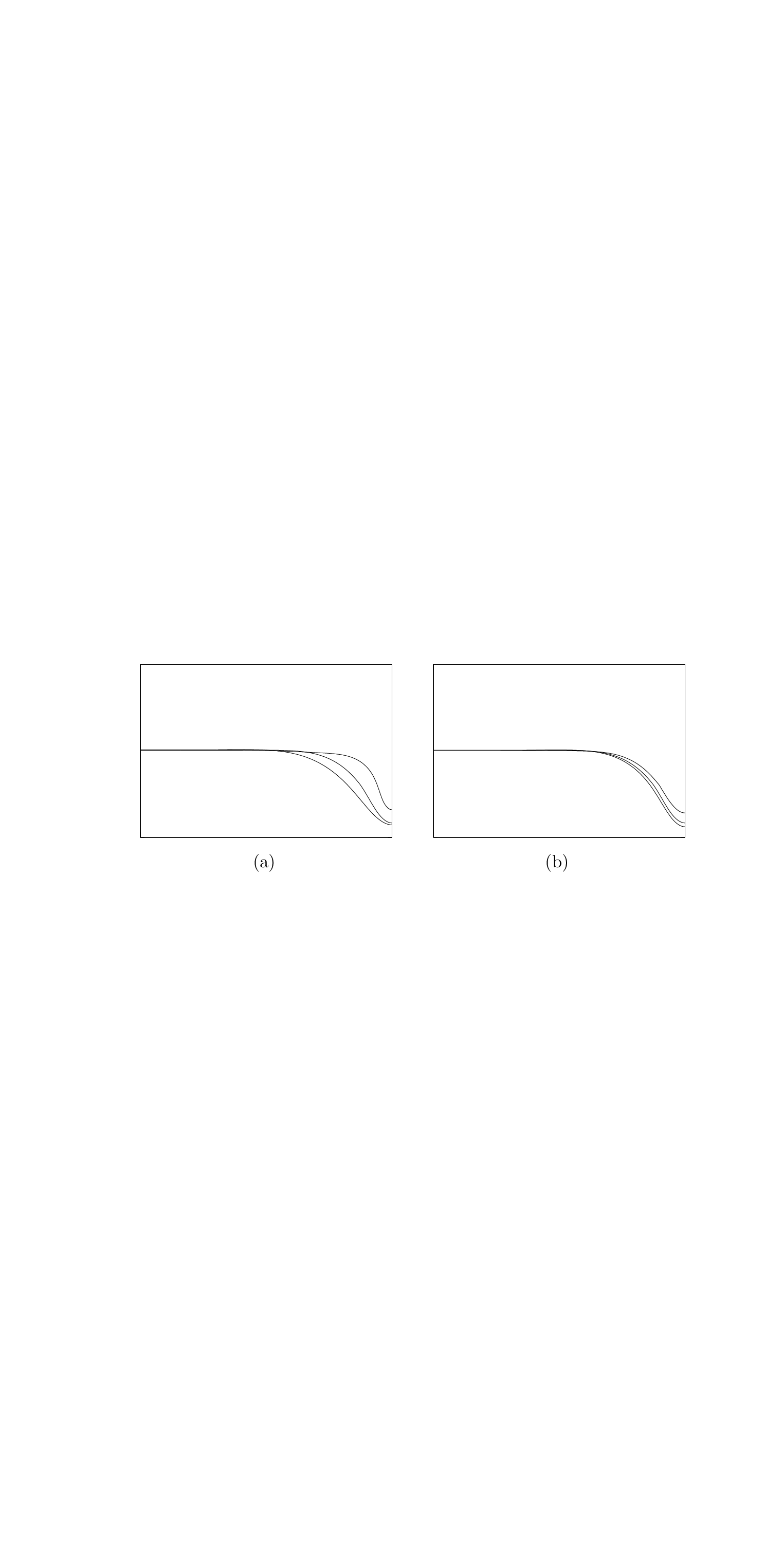}
\end{center}
\caption{Interface profiles. (a) $w_{1\infty}/b=10$, 5, and 3.33 (top to bottom profiles); $Ca=0.033$.
(b) $Ca$=0.05, 0.033, and 0.025 (top to bottom profiles);
 $w_{1\infty}/b=5$, and $w/b=10$. $\mu_1/\mu_2=1$ and $k=1$. Only half of the computational domain 
is shown.\label{fig:numerics2}}
\end{figure}

To show the effect of the flow rate of the inner phase on the narrowing of the
tip, interface profiles are shown in Figure \ref{fig:numerics2}(a) when 
varying $w_{1\infty}/b$. The main feature is that
increasing the flow rate of the inner phase results in
an abrupt change in the interface curvature, i.e.~the length 
over which the deformation of the interface takes place decreases.
If the flow rate of the inner phase is small, then the change in 
the interface curvature is more gentle. However, in this scenario, 
the Hele-Shaw approximation breaks down because $\delta < b$.
This may be the contributing factor for the different breakup mechanisms
reported in Ref.~\onlinecite{Priest2006} when changing the flow rate of the inner
phase from low to high. 

In Figure \ref{fig:numerics2}(b),
$w_{1\infty}/b$ and $w/b$ are kept constant while varying
$Ca$. Figure \ref{fig:numerics2}(b) shows that the characteristic 
length over which the abrupt change of the interface curvature
occurs is weakly dependent on $Ca$ once the capillary number is below a critical value.

\begin{figure}[t]
\centering
\includegraphics[width=0.7\textwidth]{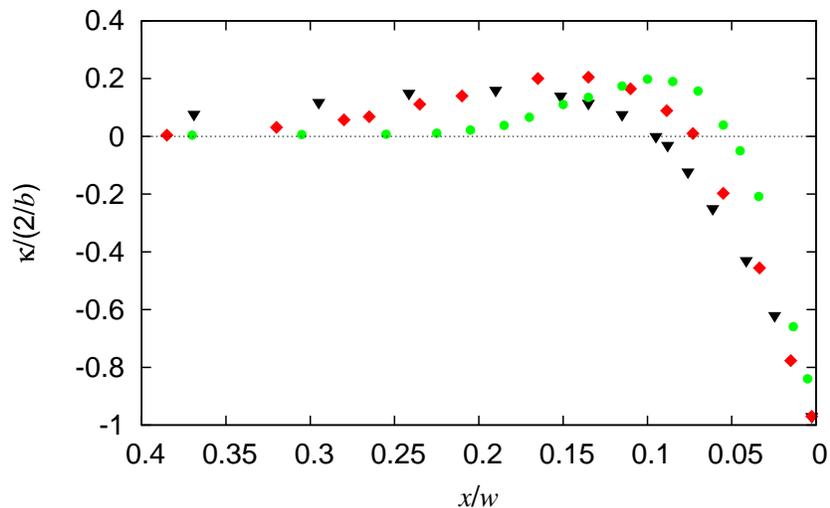}\\
\caption{Computed interface curvature normalized by $2/b$, the 
magnitude of the curvature at the outflow ($x=0$), as a function
of $x/w$ for $Ca=0.03$ ($\blacktriangledown$), 
$0.1$ ({\color{red}$\blacklozenge$}), and $0.3$ ({\color{green}$\bullet$});
the dashed line, $\kappa=0$, is only plotted to guide the eye.
$w_{1\infty}/b=5$, and $w/b=10$; $\mu_1/\mu_2=1$ and $k=1$.}
\label{fig:numerics3}
\end{figure}

To demonstrate the local change of the interface shape in the focusing region,
the computed interface curvature is shown in Figure \ref{fig:numerics3}
for $Ca=0.03$, $0.1$, and $0.3$. This shows the abrupt change 
in the narrowing region, where the  curvature changes sign from being $-2/b$
at the outflow to zero at inflow. (Note that with $p=0$ at the outflow boundary, 
we arrive at $\kappa=-2/b$ at the outlet $x=0$.) As shown, the increase in the 
capillary number results in the decrease of the length scale over which the curvature changes sign.
  
\section{Conclusions}
The formulation and implementation of a robust volume-of-fluid height-function numerical algorithm for the Hele-Shaw equations
with two immiscible liquids are presented. The components of the numerical scheme are validated with benchmark computations.
The simulation of a ribbon of fluid which co-flows with a second liquid through a Hele-Shaw cell is carried out to give a critical 
assessment of the theory of Ref.~\onlinecite{Malloggi2010} for an  estimate of the jet width at exit. The parameters
in the numerical simulations are taken from the controlled experiments in the literature  
\cite{Priest2006,Malloggi2010}.  The results  show that when the capillary number is small,  there is a region just short of the 
exit where the flowfield changes in a  complex manner, and which is not captured by  simply looking at the inflow and outflow 
fluxes. An example is the sign reversal in curvature at the exit,
 which is clearly seen in the numerical simulations. 
We also find that the effect of increasing the jet phase flow rate is to encourage the abrupt change in the interface 
curvature.  

\begin{acknowledgments}
We thank M.~Renardy, A.~Leshansky, P.~Tabeling and M-C.~Jullien for fruitful discussions. 
This research was partly supported by NSF-DMS 0907788 and NSF-DMS-1311707. 
\end{acknowledgments}

%


\end{document}